%% file: main.tex
\documentclass[]{pasj01}
\draft

\begin{document} 
\Received{2017/6/28}
\Accepted{2017/12/28}

\input{definitions}

\input{s0_title}
\maketitle
\input{s0_abstract}
\input{s1_introduction}

\input{s2_observations}

\input{s3_results}
\input{s4_discussion}

\input{s5_conclusions}

\begin{ack}
NANTEN2 is an international collaboration of 11 universities: Nagoya University, Osaka Prefecture University, University of Bonn, University of Cologne, Seoul National University, University of Chile, University of New South Wales, Macquarie University, University of Sydney, University of Adelaide, and University of ETH Zurich. The Mopra radio telescope is part of the Australia Telescope National Facility. The University of New South Wales, the University of Adelaide, and the National Astronomical Observatory of Japan (NAOJ) Chile Observatory supported operations. The ASTE telescope is operated by NAOJ. This work was financially supported by Grants-in-Aid for Scientific Research (KAKENHI) of the Japanese society for the Promotion of Science (JSPS, grant No. 15H05694).  Finally, we are grateful to the referee for his/her thoughtful comments. 
\end{ack}

\input{s9_references}
\input{s9_figures}

\input{s9_tables}

\end{document}

%% file: definitions.tex
\def\Herschel{{\it Herschel}}
\def\Spitzer{{\it Spitzer}}
\def\WISE{{\it WISE}}
\def\AKARI{{\it AKARI}}

\def\um{$\mu \rm{m}$}

\def\Xco{$X_{\rm CO}$}
\def\Xunit{$\rm{cm}^{-2}$ $\rm{K}^{-1}$ $\rm{km}^{-1}$ $\rm{s}$}
\def\cmcm{$\rm{cm}^{-2}$}
\def\cmcmcm{$\rm{cm}^{-3}$}
\def\kms{$\rm{km}$ $\rm{s}^{-1}$}
\def\vlsr{$v_{\rm LSR}$}
\def\degree{$^{\circ}$}
\def\Lsun{$L_{\solar}$}
\def\Msun{$M_{\solar}$}
\def\Msunyr{$M_{\solar}$ $\rm{yr}^{-1}$ }

\def\NH{$N(\rm{H_{2}})$}

\def\HII{H \emissiontype{II}}
\def\OIII{O \emissiontype{III}}
\def\SII{S \emissiontype{II}}

\def\lb{($l$, $b$)}
\def\lbeq{($l$, $b$)$=$}
\def\lbsim{($l$, $b$)$\sim$}
\def\radec{($\alpha_{\rm J2000}$, $\delta_{\rm J2000}$)}
\def\radeceq{($\alpha_{\rm J2000}$, $\delta_{\rm J2000}$)$=$}
\def\radecsim{($\alpha_{\rm J2000}$, $\delta_{\rm J2000}$)$\sim$}

\def\COa{\atom{C}{}{12}\atom{O}{}{}}
\def\COb{\atom{C}{}{13}\atom{O}{}{}}
\def\COc{\atom{C}{}{}\atom{O}{}{18}}

\def\Jeq{{\it J}$=$}
\def\Ja{({\it J}$=1$--$0$)}
\def\Jb{({\it J}$=2$--$1$)}
\def\Jc{({\it J}$=3$--$2$)}

%% file: s0_title.tex
\title{Molecular clouds toward three Spitzer bubbles S116, S117 and S118: Evidence for a cloud-cloud collision which formed the three \HII \ regions and a 10-pc scale molecular cavity}

\author{Yasuo \textsc{Fukui}\altaffilmark{1,2,*}}
\author{Akio \textsc{ohama}\altaffilmark{1}}
\author{Mikito \textsc{Kohno}\altaffilmark{1}}
\author{Kazufumi \textsc{Torii}\altaffilmark{3}}
\author{Shinji \textsc{Fujita}\altaffilmark{1}}
\author{Yusuke \textsc{Hattori}\altaffilmark{1}}
\author{Atsushi \textsc{nishimura}\altaffilmark{1}}
\author{Hiroaki \textsc{yamamoto}\altaffilmark{1}}
\author{Kengo \textsc{tachihara}\altaffilmark{1}}

\altaffiltext{1}{Department of Physics, Nagoya University, Furo-cho, Chikusa-ku, Nagoya, Aichi 464-8601, Japan}
\altaffiltext{2}{Institute for Advanced Research, Nagoya University, Furo-cho, Chikusa-ku, Nagoya 464-8601, Japan}
\altaffiltext{3}{Nobeyama Radio Observatory, 462-2 Nobeyama, Minamimaki Minamisaku, Nagano, 384-1305, Japan}
\email{fukui@a.phys.nagoya-u.ac.jp}


%% file: s0_abstract.tex
\begin{abstract}
We carried out a molecular line study toward the three Spitzer bubbles S116, S117 and S118 which show active formation of high-mass stars. 
We found molecular gas consisting of two components with velocity difference of {$\sim 5$ \kms}. 
One of them, the small cloud, has typical velocity of {$-63$ \kms} \ and the other, the large cloud, has that of $-58$ \kms. 
The large cloud has a nearly circular intensity depression whose size is similar to the small cloud. 
We present an interpretation that the cavity was created by a collision between the two clouds and the collision compressed the gas into a dense layer elongated along the western rim of the small cloud.  
In this scenario, the O stars including those in the three Spitzer bubbles were formed in the interface layer compressed by the collision. 
By assuming that the relative motion of the clouds has a tilt of \timeform{45D} to the line of sight, we estimate that the collision continued over the last 1 Myr at relative velocity of $\sim$10 \kms. 
In the S116--117--118 system the \HII \ regions are located outside of the cavity.
This morphology is ascribed to the density-bound distribution of the large cloud which made the \HII \ regions more easily expand toward the outer part of the large cloud than inside of the cavity. 
The present case proves that a cloud-cloud collision creates a cavity without an action of O star feedback, and suggests that the collision-compressed layer is highly filamentary.
\end{abstract}

%% file: s1_introduction.tex
\section{Introduction}
O stars have a mass range from 20 to 300 \Msun \ in the Local Group \citep{wal02}. 
They influence the interstellar space considerably by dynamically disturbing the interstellar medium (ISM) over their whole lifetime and by enriching the heavy elements at the end of their lives via supernova explosions.
These actions have a significant influence on the galactic evolution. 
It is therefore of crucial importance to better understand the formation mechanism of O stars.
The mass accretion rate required for O star formation is as high as 10$^{-4}$ \Msunyr; otherwise the high stellar radiation pressure halts the mass accretion and higher-mass stars do not grow in mass to a mature O star (e.g., \cite{wol87, zin07}).
 
It is often discussed that dense and massive clouds like infrared dark clouds are a plausible site of high-mass star formation. 
This seems to be a natural scenario, whereas we have often no \HII \ regions associated with infrared dark clouds; regions where O stars are formed unambiguously are \HII \ regions which are ionized by ultraviolet photons emitted by O or early B stars.
\HII \ regions in the Galaxy show a variety of morphology and evolutionary stages, from ultra-compact \HII \ regions to extended \HII \ regions (e.g., for review \cite{chu02}), and more careful studies of \HII \ regions have a potential to shed a new light on O star formation.
In particular, a detailed study of molecular gas, the raw material to form stars, is still not thoroughly made in \HII \ regions due to a limited angular resolution or a low dynamic range in molecular line observations in spite of the previous efforts on molecular observations toward \HII \   regions (e.g., Andreson et al. 2009).

Spitzer bubbles are seen at 8 \um \ with a ring-like shape and harbor usually an O star and an \HII \ region in the ring. 
In total 322 bubbles are cataloged in Galactic longitude from \timeform{-65D} \ to \timeform{65D} and in Galactic latitude from \timeform{-1.5D} to \timeform{1.5D} except the Galactic center ($l< \pm$ \timeform{10D}) \citep{chu06}.
It is usually discussed that the ultraviolet photons and stellar winds accelerate and compress the surrounding gas to form an expanding bubble and trigger second-generation star formation (the collect-collapse scenario) \citep{deh05}.
A CO survey of Spitzer bubbles was made by \citet{bea10} who observed 43 Spitzer bubbles with JCMT in the CO \Jeq 3--2 transition.
Their results indicate that the gas is ring-like and flattened with no significant sign of expansion as opposed to the wind-blown scenario, raising a puzzle which is not immediately explained by the conventional picture. 
 
\citet{tor15}, recently, presented a novel scenario that the bubble in a most typical Spitzer bubble RCW120 \citep{zav07, deh09} was formed by cloud-cloud collision via employing the \citet{hab92} model, which numerical simulated head-on collision between small and large spherical clouds.
These simulations showed that such uneven collision can create a cavity in the large cloud with a size of the small cloud and the compressed layer between them become dense and self-gravitating to trigger star formation. 
If the formed star is able to ionize the surroundings, i.e., the inside of the cavity, the model provides an alternative to explain the ring-like shape of RCW120 inside of which is ionized \citep{tor15}. 
\citet{tor15} did not find an observational signature of expansion of the bubble in the molecular gas, which lead to the cloud-cloud collision scenario.
RCW120 has two velocity molecular components, one of which shows a good correspondence with the bubble, and the other associated with the opening of the bubble. 
These properties are consistent with the results by \citet{bea10}.
The two clouds show enhanced temperature toward the ring in spite of their large velocity separation $\sim20$ \kms, which provides robust verification of their physical association with RCW120. 
Since the velocity separation is too large to be bound by the cloud gravity, \citet{tor15}  concluded that RCW120 is a case of cloud-cloud collision which triggered formation of a single O7 star. 
It is noteworthy that the new model offers an explanation of the elongated horseshoe morphology of the bubble along with the off-center position of the exciting star close to the bubble bottom; the simple stellar-wind bubble does not offer a natural explanation for such asymmetry. 

Figure \ref{fig:RGB} shows the region of three Spitzer bubbles S116, S117, and S118.
Distance of S116 is estimated kinematically to be $5.9 \pm 0.9$ kpc \citep{chu06}.
The green color and red color in Figure \ref{fig:RGB} show 8 \um \ and 24 \um \ images, respectively.
The 8 \um \ radiation traces the PAH emission and the 24 \um \ radiation the heated dust by the ultraviolet radiation of the O stars.
The bubbles are not perfectly circular and are elongated significantly.
The Spitzer bubbles are often isolated, and such a case with three lined-up bubbles are rare.

In order to have a better understanding of formation of multiple Spitzer bubbles, we undertook new CO observations of molecular gas toward S116, S117, and S118 by using three mm/sub-mm telescopes NANTEN 2, ASTE, and the Mopra 22-m telescope in the CO rotational transitions.
Section 2 gives description of observations and Section 3 presents observational results. 
Section 4 gives discussion of the results and tests a cloud-cloud collision scenario, and Section 5 concludes the paper.

%% file: s2_observations.tex
\section{Observations}
\subsection{\COa \ \Jeq 1--0 with NANTEN2}
We performed \COa \ \Jeq 1--0 observations with the NANTEN2 4-m millimeter sub-millimeter radio telescope \citep{miz04}.
The observations were carried out in October 2011.
The front end was a 4 K cooled SIS mixer receiver.
The typical system noise temperature including the atmosphere was $\sim$250 K in the double-side band (DSB) during observations. 
The backend was a digital-Fourier transform spectrometer (DFS) with 16384 channels of 1 GHz bandwidth, which corresponds to velocity coverage and resolution of $\sim$2600 \kms \ and 0.16 \kms, respectively. 
We used the on-the-fly (OTF) mapping mode to cover an area of \timeform{1D}$\times$ \timeform{1D}.
The pointing accuracy was confirmed to be better than \timeform{15''} with daily observations toward the Sun.
The absolute intensity was calibrated by observing Orion KL everyday.
The final half-power beam width (HPBW) is $180''$. 
The final root-mean-square (rms) noise fluctuation of the data was $\sim$ 1.0 K/ch at a velocity resolution of 0.16 km s$^{-1}$.

\subsection{\COa \ \Jeq 1--0 and \COb \ \Jeq 1--0 with Mopra}
Detailed \COa \ \Jeq 1--0 and \COb \ \Jeq 1--0 distributions around the Spitzer bubbles S116, S117, and S118 were obtained by using the 22-m ATNF (Australia Telescope National Facility) Mopra mm telescope at a high angular resolution of \timeform{33''} in two periods June-October 2013 and July-August 2014. 
For the field shown in Figure 1 by a dashed box we simultaneously observed \COa \ \Jeq 1--0 and \COb \ \Jeq 1--0 in the OTF mode with a unit field of \timeform{4'} $\times$ \timeform{4'}.
The typical system noise temperature including the atmosphere was between 400 K and 600 K in the single-side band (SSB) during observations. 
The Mopra backend system ``MOPS" which provides 4096 channels across 137.5 MHz in each of the two orthogonal polarizations was used in the observations. 
The velocity resolution was 0.088 \kms \ and the velocity coverage was 360 \kms \ at 115 GHz. 
The pointing accuracy was checked every 90 minutes to keep within \timeform{4''} with observations of 86 GHz SiO masers.
The absolute intensity was calibrated by comparing with CO \Jeq 1--0 data observed with NANTEN2.
The obtained data were smoothed to a HPBW of \timeform{40''} with a 2D Gaussian function and to a 0.6 \kms \ velocity resolution. 
The final data cube had a rms noise fluctuation of $\sim$ 0.6 K/ch at a velocity resolution of 0.088 km s$^{-1}$.

\subsection{\COa \ \Jeq 3--2  with ASTE}
Observations of the \COa \ \Jeq 3--2 emission were performed by using the ASTE 10-m telescope located in Chile in three periods September 2013, June 2014, and November 2015.
The waveguide-type sideband-separating SIS mixer receiver for the single sideband (SSB) "CAT345'' having system noise temperature of 300 K and the digital spectrometer "MAC'' with the narrow-band mode providing 128 MHz bandwidth and 0.125 MHz resolution, which corresponds to 450 \kms \ velocity coverage and 0.43 \kms \ velocity resolution at 345 GHz, were used. 
The observations were made with the OTF mode at a grid spacing of \timeform{7.5''}, and the HPBW was \timeform{24''} at the \COa \ \Jeq 3--2 frequency.
The Observed area is the same as that of the Mopra observations.
The pointing accuracy was checked every 90 minutes to keep within \timeform{2''} with observations of RAFGL 4202 \radeceq (\timeform{14h52m23.82s}, \timeform{-62D04'19.2"}).
The absolute intensity was calibrated by observing W44 and IRC+10216 every 90 minutes.
The obtained data was smoothed to a HPBW of \timeform{22''} with a 2D Gaussian function and to a 1 \kms \ velocity resolution. 
The final data cube had a rms noise fluctuation of $\sim$ 0.38 K/ch at a velocity resolution of 0.43 km s$^{-1}$.

%% file: s3_results.tex
\section{Results}
\subsection{CO distributions}
Figure \ref{fig:RGB} shows an infrared image of the present region taken with Spitzer \citep{ben03,car09}.
The three Spitzer bubbles S116, S117 and S118 are distributed over 20 pc in the north-south direction.
We find nearly ten smaller \HII \ regions, which are bright warm dust traced by the Spitzer 24 $\mu$m emission. 
The area indicated by the dashed-line box was observed with Mopra and ASTE, while the whole area was mapped with NANTEN2. 
Figures \ref{fig:nanten}a and \ref{fig:nanten}b show a large-scale view of the \COa \ \Jeq 1--0 emission observed with NANTEN2.
We find two clouds whose distributions are distinctly different.
The $-63$ \kms cloud has a peak at \lbeq (\timeform{314.22D}, \timeform{0.33D}).
The $-58$ \kms cloud is extended along the plane and has a peak at \lbeq (\timeform{314.21D}, \timeform{0.25D}) and an intensity depression at \lbeq (\timeform{314.3D}, \timeform{0.36D}) in addition to several intensity peaks surrounding. 
We shall hereafter call the $-63$ \kms cloud and the $-58$ \kms cloud the small cloud and the large cloud, respectively, because of their sizes.  The small cloud has a sharp intensity decrease to every direction. The large cloud shows a sharp intensity gradient toward the west.

Figure \ref{fig:chmap} shows velocity channel distributions every 1.3 \kms \ in the \COa \ \Jeq 1--0 emission taken with the Mopra telescope, which indicates that the small cloud becomes large in size with velocity increase from $-69.3$ \kms \ to $-62.7$ \kms.
We also find that the CO distribution is extremely filamentary in a velocity range from $-66.6$ \kms \ to $-61.3$ \kms \ particularly toward the small cloud. The filamentary structure is described into more detail in section \ref{sec:ratio}.

Figure \ref{fig:CO_3x2} shows detailed distribution of \COa \ \Jeq 1--0, \COb \ \Jeq 1-0 and \COa \ \Jeq 3--2 images of the small cloud (Figure \ref{fig:CO_3x2}a, 4b, and 4c) and the large cloud (Figure \ref{fig:CO_3x2}d, 4e, and 4f) overlaid with infrared contours. 
The bubbles S116 and S118 delineate the northern and southern boundary of the small cloud, and S117 is located toward the small cloud.
In the large cloud, the cavity is clearly seen with a sharp nearly circular boundary. 
We find another intensity depression in the north at \lbeq (\timeform{314.27D} --\timeform{314.32D}, \timeform{0.42D} --\timeform{0.44D}).
The western edge of the large cloud shows a distribution similar to the small cloud at $b$ = \timeform{0.3D} --\timeform{0.45D}, and S116 and S118 are located also along the edge of the large cloud. 
In Figures \ref{fig:CO_3x2}b and \ref{fig:CO_3x2}e the distribution of the $^{13}$CO $J=$1--0 emission, which is likely optically thin, shows correspondence with the intense part of the $^{12}$CO emission. 
We find that the effect of self-absorption is small in this region from the similarity between the $^{12}$CO and $^{13}$CO distributions.
The physical parameters of the two clouds are as follows. 
The size of the cavity hole is $\sim$5 pc in radius for an assumed distance of 5.9 kpc.
The mass and peak column density of the small cloud and the large cloud in the area shown in Figure \ref{fig:nanten} are ($2.0 \times 10^4$ \Msun \ , $0.7 \times 10^5$ \Msun) and ($1 \times 10^{22}$ cm$^{-2}$, $2 \times 10^{22}$ cm$^{-2}$), respectively, where an X(CO) factor of $1.0 \times 10^{20}$ \Xunit \ \citep{oka17} was assumed.

\subsection{Radio continuum distribution and the properties of the O stars}
In Figure \ref{fig:radio} the ASTE $^{12}$CO $J=$3--2 images of the two components are superposed on the SUMMS 843MHz radio continuum distribution \citep{boc99, mau03} (Figures \ref{fig:radio}a and \ref{fig:radio}b) and the Spitzer image (Figures \ref{fig:radio}c and \ref{fig:radio}d).
We find that the radio continuum distribution coincides well with the Spitzer bubbles.
S117 is located toward the center of the small cloud, and it fits the intensity depression in the large cloud (Figure \ref{fig:radio}c and \ref{fig:radio}d).
The heavy obscuration does not allow us to directly observe the exciting stars, and instead we used the radio continuum radiation as a measure of the stellar ultraviolet radiation.
We estimate the spectral types of the exciting O stars in each \HII \ region from the radio flux by using the relationship given by \citet{pan73} and \citet{kur94} as shown in Table 1, where the exciting star is assumed to be a single star in each \HII \ region. 
The ultraviolet photon flux was used to estimate stellar spectral types and corresponding stellar mass for ZAMS (zero-age main sequence).
As a result, the spectral types are estimated as follows; O6--06.5 for S116, O9 for S117, O7.5--O8 for S118, and the mass of luminosity class V stars are inferred to be 30 \Msun, 20 \Msun, and 23 \Msun, respectively \citep{mar05}.

\subsection{The intensity ratio of the \COa \ \Jeq 3--2 to the \COa \ \Jeq 1--0}\label{sec:ratio}
Figures \ref{fig:ratio_3x2}a and \ref{fig:ratio_3x2}d show the intensity ratio of the \COa \ \Jeq 3--2 to the \COa \ \Jeq 1--0 ($R_{3-2/1-0}$) for the two clouds.
$R_{3-2/1-0}$ is mainly affected by the \COa \ \Jeq 3--2 distribution. 
The typical $R_{3-2/1-0}$ of molecular clouds in our galaxy without an extra heat source is $R_{3-2/1-0}$ $\sim 0.4$ \citep{fuk16}.
$R_{3-2/1-0}$ is enhanced to 1.0--1.4 outside of the cavity toward S116 and S118, while $R_{3-2/1-0}$ is 0.6--1.0 around the cavity.
The enhanced ratio toward S116 and S118 suggests that the stellar radiation is heatig the gas. 
Red and blue lines in Figures \ref{fig:ratio_3x2}b and \ref{fig:ratio_3x2}e represent the filamentary structures of $R_{3-2/1-0}$, where red and blue colors show that the filament is overlapping with the \HII \ regions (red), or not (blue). 
These filamentary structures were identified by eye with a criteria of $R_{3-2/1-0}>0.9$ and length $>\sim2$\,pc. 
The filamentary structures have a typical size of $\sim1 \times 3$\,pc. 
Multiple filamentary structures are seen not only toward the \HII \ region but also in the cavity without \HII \ regions.

%% file: s4_discussion.tex
\section{Discussion}
We have carried out large scale CO observations with NANTEN2 in the region of the three Spitzer bubbles S116--S117--S118, which includes at least several O stars in addition to a few tens of low mass stars as Akari point sources (Figure \ref{fig:RGB}).
We made follow up high resolution observations with ASTE and Mopra telescopes.
We found that the molecular gas comprises at least two velocity components of different morphology; the $-63$ \kms cloud (the small cloud) and the $-58$ \kms cloud (the large cloud), which are extended along the plane over 40 pc if the weak extended CO emission is included in Figure \ref{fig:nanten}.
The large cloud has an intensity peak and an intensity depression of $\sim 10$ pc in size. 
Higher resolution observations with Mopra and ASTE show that the small cloud grows in size with the increase of velocity (Figure \ref{fig:chmap}).

\subsection{A cloud-cloud collision scenario in S116--S117--S118}
In order to explain the velocity distribution we propose a hypothesis that cloud-cloud collision took place between the two clouds and that the small cloud pushed the surface of the large cloud to produce the cavity in the large cloud.  
Figure \ref{fig:takahira_chmap}a prsents a schematic view of the collision seen from a direction perpendicular to the cloud relative motion.
In the plane of Figure \ref{fig:takahira_chmap}a, $\theta$ is an angle of the line of sight to the relative motion of the cloud. 
The small cloud was separated in the upper left corner of Figure \ref{fig:takahira_chmap}a prior to collision, and moved along a straight line toward the large cloud. 
They collided with each other and the small cloud created a cavity in the large cloud. 
The layer between the two clouds are compressed to form an interface layer, which is denoted by dark blue in front of the small cloud in Figure \ref{fig:takahira_chmap}a.
The two clouds observed on the sky are divided into three sections A, B, and C as shown in Figure \ref{fig:takahira_chmap}a. 
A shows the large cloud and the cavity, B the interface layer, the small cloud, and the large cloud with the cavity, and C shows part of the large cloud prior to collision.

In order to gain an insight into observed cloud properties, we describe the physical states of colliding clouds by using the hydrodynamical numerical simulations of \citet{tak14}.
The simulations deal with head-on collision between a small cloud and a large cloud, which are idealized to be spherically symmetric.
We adopt a model listed in Table 2 for discussion; the radius of the small cloud is 3.5 pc and that of the large cloud 7.2 pc.
The two are currently colliding at 7 \kms \ and have internal turbulence in the order of 1--2 \kms \ with highly inhomogeneous density distribution.
For more details see \citet{tak14}.
The cloud parameters do not correspond exactly to the present cloud although the difference does not critically affect a qualitative comparison below.

We assume $\theta = 45^{\circ}$ in the following and an epoch of 1.6 Myr after the onset of the collision, where cloud signatures typical to collision are seen.
The assumption on $\theta$ is not so critical as long as $\theta$ is not very close to $0^{\circ}$ or $90^{\circ}$.
The small cloud is producing a cavity in the large cloud by the collisional interaction.
The interface layer of the two clouds has enhanced density by collision, where the internal turbulence and the momentum exchange between the clouds mix the gas distribution in space and velocity.
The gas in the two clouds is continuously merging into the layer during the collision.
Figures \ref{fig:takahira_chmap}b--\ref{fig:takahira_chmap}i show the velocity channel distributions every 0.5 \kms.
In cloud-cloud collision it might be expected to see two distinct clouds of a narrow line width, whereas the actual distribution in the simulations present a merged single cloud which is continuous in velocity and space.
The small cloud is seen at a velocity range from $-5.1$ to $-2.2$ \kms, and the large cloud from --$1.2$ to $1.7$ \kms.
The velocity range from $-2.2$ to $-1.2$ \kms \ corresponds to the interface velocity layer which was created by merging.
Note that the velocity ranges of each panel in Figure \ref{fig:takahira_chmap} do not exactly fit the velocity ranges of the two clouds due to the mixing in velocity.
We find that the small cloud becomes large with the increase of velocity by merging of the large cloud as is consistent with the observations.
The cavity is produced in the large cloud by the small cloud, while the boundary of the cavity is less clear in the model than the observations, reflecting that turbulence is more enhanced in the model than in the observed cloud.

In S116-117-118, there is a displacement of $\sim10$ pc in the sky between the small cloud and the cavity (Figure \ref{fig:chmap}).
We interpret that the displacement is caused by projection of a tilt of the cloud relative motion to the line of sight.
The collision velocity is estimated to be $\sim7$ \kms \ if we assume tentatively $\theta = 45$ deg, an angle between the relative motion to the line of sight.

Figure \ref{fig:pv} shows a comparison in a position-velocity diagram taken in the direction of the relative motion of the two clouds.
Figure \ref{fig:pv}a is an overlay of the small and large clouds taken from Figure \ref{fig:takahira_chmap}e and Figure \ref{fig:takahira_chmap}h. 
Figure \ref{fig:pv}a shows distributions of the two clouds. 
Figure \ref{fig:pv}b shows a synthetic position-velocity-diagram produced from the same numerical simulations in the plane of the two cloud centers.
The clouds as a whole show a ``V-shape'' as traced by the solid lines, and the main peak is found at X$=$5--7 pc and V$=-4$--$-1$ \kms in Figure \ref{fig:pv}b.
This peak is formed by a combination of the small cloud and the interface layer which merged together.
There is an intensity depression at X$=$4--6 pc and V$=-2$--0 \kms, which correspond to the blue-shifted part of the cavity in the large cloud. 
Figure \ref{fig:pv}c is the observed two velocity components and Figure \ref{fig:pv}d the observed position-velocity diagram.
The ''V-shape'' typical to collision is recognized in the Figure \ref{fig:pv}d.
Correspondence is seen between the model and the observations qualitatively; 
the synthetic observations reproduce the cavity at  $l=$\timeform{314.3D} \ and \vlsr  $=-58$ \kms \ and the small cloud at $l=$\timeform{314.2D} \ and \vlsr $=-64$ \kms, except for the sharp cut of the large cloud at  $l=$\timeform{314.2D} \ which is not taken into account in the model.
This correspondence lends support for applying the cloud-cloud collision model to the S116-S117-S118 system.

The interface layer is strongly compressed by collision, and the O stars ionizing the three Spitzer bubbles S116-117-118 were formed in the layer due to gravitational instability.
This explains the distribution of the bubbles along the western surface of the small cloud which is interacting with the large cloud. 
The timescale of the collision is roughly estimated to be $\sim 1.3$ Myr ($=10$ pc / $7$ \kms).
An O stars of 30 \Msun \ is formed in $10^{5}$ yr within the timescale for a mass accretion rate $3 \times 10^{-4}$ \Msunyr, which is adopted from a typical value in the compressed layer of cloud-cloud collision in magnetohydrodynamical numerical simulations \citep{ino13}.
This satisfies the mass accretion rate required to form O stars by overcoming the stellar radiation feedback \citep{wol87}. 

The collisional compression is possibly extended over 40 pc beyond the size of the cavity in the north-south direction vertical to the projected motion of the two clouds.
The layer is observed as a north-south elongated molecular ridge at $l=314.2$ from 0.1 to 0.6 in $b$, and we find a possible sign of further triggered formation of lower mass stars along the compressed layer found as at least several compact infrared sources in Figure \ref{fig:RGB}.
So, it is possible that triggering is extensive in space, while O star formation is probably limited to the region of high molecular column density $1$--$2\times 10^{22}\,{\rm cm^{-2}}$.

\subsection{Comparison with RCW120}
In S116-117-118, the \HII \ regions are located outside the collision-created molecular cavity, although in RCW120 the \HII \ region is located inside the cavity. 
In the both cases the collision formed the interface layer where O star(s) form, and the morphological difference between the two cases are to be explained.
The schematic drawings of the two colliding clouds in position-velocity diagram of the two cases, RCW 120 and S116-117-118, are shown in Figure \ref{fig:schematic_pv}. 
RCW120 \citet{tor15} \ presented a cloud-cloud collision model in order to explain the formation of the O star within the bubble.
The diameter of the RCW120 bubble is $\sim3$ pc, whereas that of the S116-S117-S118 cavity is $\sim10$ pc. 
Therefore, the volume of the S116-S117-S118 cavity is larger than the RCW120 cavity by a factor of $\sim30$ if a spherical cavity is assumed, and the molecular mass inside the cavity is significantly larger in S116-S117-S118 than in RCW120. 
This offers a possible explanation for that the larger number of O stars in the present case. 
Figure \ref{fig:schematic_star} shows the schematic images of cloud-cloud collision in RCW 120 and S116--S117--S118. 
According to the \citet{hab92} model of cloud-cloud collision, we expect that two clouds, one of which is delineating a Spitzer bubble and the other localized inside the bubble, exhibit complementary distribution in the early phase after the collision (Phase 1 in Figure \ref{fig:schematic_star}).
Later, the small cloud is destructed by ionization due to the formed star and by collisional merging into the interface layer in RCW120. 
The difference between the two cases is the location of the O stars. In RCW120 the O star is inside the cavity (Figure \ref{fig:schematic_star}a Phase 3), whereas in S116--S117--S118 the O stars are located outside of the cavity (Figure \ref{fig:schematic_star}b Phase 3). 
In S116-117-118, we suggest that the collision happened by chance close to the edge of the large cloud. 
This situation made the shocked layer exposed to the outside of the large cloud where density drops suddenly. 
We infer that this ad foc geometry in S116-117-118 caused the O star formation outside the collision-created molecular cavity. 

\subsection{The physical properties of the collisional interface layer}
The S116--S117--S118 cavity demonstrates clearly the role of cloud-cloud collision in creating a cavity, which shows a well-ordered nearly circular boundary. 
Because no O star exists inside the cavity, there is no room for the cavity to be produced by an O star.
In  RCW120 the inside wall of the cavity is partially ionized and the physical conditions are affected by the ionization, implying that the collisional interface do not keep the conditions just after the collision. 
The present case is different because it provides physical conditions in the shock-compressed layer unaffected by the ionization. 
Although S117 seems to lie toward the peak of the small cloud, S117 is not affecting the small cloud and the inside of the cavity as shown by no enhanced $R_{3-2/1-0}$ ratio toward S117 in Figure \ref{fig:ratio_3x2}d.
In this context, we pay attention to that the interface layer toward the small cloud and the cavity exhibit highly filamentary distribution which is obvious in the $R_{3-2/1-0}$ distribution in Figure \ref{fig:ratio_3x2}a. 
\citet{ino17} showed that molecular filaments are formed in the shocked interface in a cloud-cloud collision, and similar filament formation is seen in the simulations by \citet{tak14} and \citet{bal17}. 
The width and length of the filaments in these simulations deserve a further detailed comparison with observations in order to better understand the physical processes in collision.

Figures \ref{fig:ratio_3x2}c and \ref{fig:ratio_3x2}f show the distribution of \COb \ \Jeq 1-0 images of the small cloud and the large cloud overlaid on the filamentary distribution of $R_{3-2/1-0}$. 
The high \COb \ \Jeq 1-0 intensity part and the $R_{3-2/1-0}$ filamentary structures in the collision interface are not corresponding with each other, but the high $R_{3-2/1-0}$ regions rather correspond to the edge of the \COb \ \Jeq 1-0 intensity peaks. 
Since there is no \HII \ region toward the cavity, the high $R_{3-2/1-0}$ is possibly due to heating by the collisional shock but not by the enhancement of density or irradiation by ultraviolet photons of O stars. 
In the density regime concerned molecular cooling time scale is $\sim 10^4$ yr and  collisional shock heating may by responsible for the high line intensity ratio in such a small timescale. 
S116--S117--S118 may be a rare case where heating by ultraviolet radiation is not important in a collision-produced cavity, allowing us to test isolate contribution of the shock heating only.

%% file: s5_conclusions.tex
\section{Conclusions}
We carried out a molecular line study toward the three Spitzer bubbles S116, S117 and S118. 
The region is associated with nearly ten smaller \HII \ regions, indicating active formation of high-mass stars over a length of 50 pc although the region did not attract much attention until now.
The detailed molecular data in the present work lead to the following conclusions which offer a novel insight into the formation of the Spitzer bubbles; 

\begin{enumerate}
\item 
The molecular clouds in the region of the three Spitzer bubbles S116--S117--S118 include two velocity components; one of them, the small cloud, has  $-63$ \kms \ and the other, the large cloud, has $-58$ \kms, while the two clouds appear to be continuously distributed in a position-velocity diagram. 
The large cloud has a cavity, an intensity depression, which is apparently correlated with the morphology of the small cloud. 
The two of the Spitzer bubbles S116 and S118 and additional small \HII \ regions are distributed along the northwestern and southwestern edges of the small cloud, and the other S117 toward the peak of the small cloud.

\item 
We present an interpretation that the cavity was created by a collision between the two clouds  $\sim$1 Myr ago and the collision compressed the gas into a dense layer elongated in the north-south direction over an extent of $\sim$20 pc at a kinematic distance of 5.9 kpc. 
In the compressed layer produced by the collision the O stars exciting the \HII \ regions including the three Spitzer bubbles were formed. 
We show that a position-velocity diagram including the small cloud and the cavity exhibits a pattern characteristic to cloud-cloud collision simulated numerically by \citet{tak14}. 
By assuming that the relative motion of the clouds has $45^{\circ}$ to the line of sight, we estimate that the collision velocity and the collision timescale to be $\sim 7$ \kms \ and $\sim 1$ Myr, respectively. 

\item 
The morphology is different from the collision-induced star formation in RCW120 where the \HII \ region was formed within the cavity created by the collision. 
We suggest the difference is due to the density-bound distribution of the large cloud in S116-117-118, which made the \HII \ regions expand toward the less dense outer part of the cloud rather than the denser inside. 
The present case is important in two ways.
One is that it demonstrates unambiguously formation of a molecular cavity by cloud-cloud collision without O stars. 
Another is that it allows us to watch details of the collision compressed layer with no influence of ultraviolet photons of O stars; in particular, the highly filamentary distribution is seen as is consistent with the numerical simulations of cloud-cloud collision that predicted filamentary distributions in the shocked gas \citep{ino13, tak14, bal17, ino17}.
\end{enumerate}

To summarize, we conclude that the three \HII \ regions are ionized by the O stars formed by triggering  in the cloud-cloud collision.
This is a case where \HII \ regions expanded outside the collision-created cavity.
The collision time scale is estimated to be $\sim$1 Myr.
A mass accretion rate of $3 \times 10^{-4}$ \Msunyr \ based on MHD simulations of cloud-cloud collision \citep{ino13} explains O star formation in $10^{5}$ yr, significantly smaller than the collision duration, which is consistent with the O star formation in the late phase of the collision.

%% file: s9_figures.tex
\begin{figure}
 \begin{center}
  \includegraphics[width=16cm]{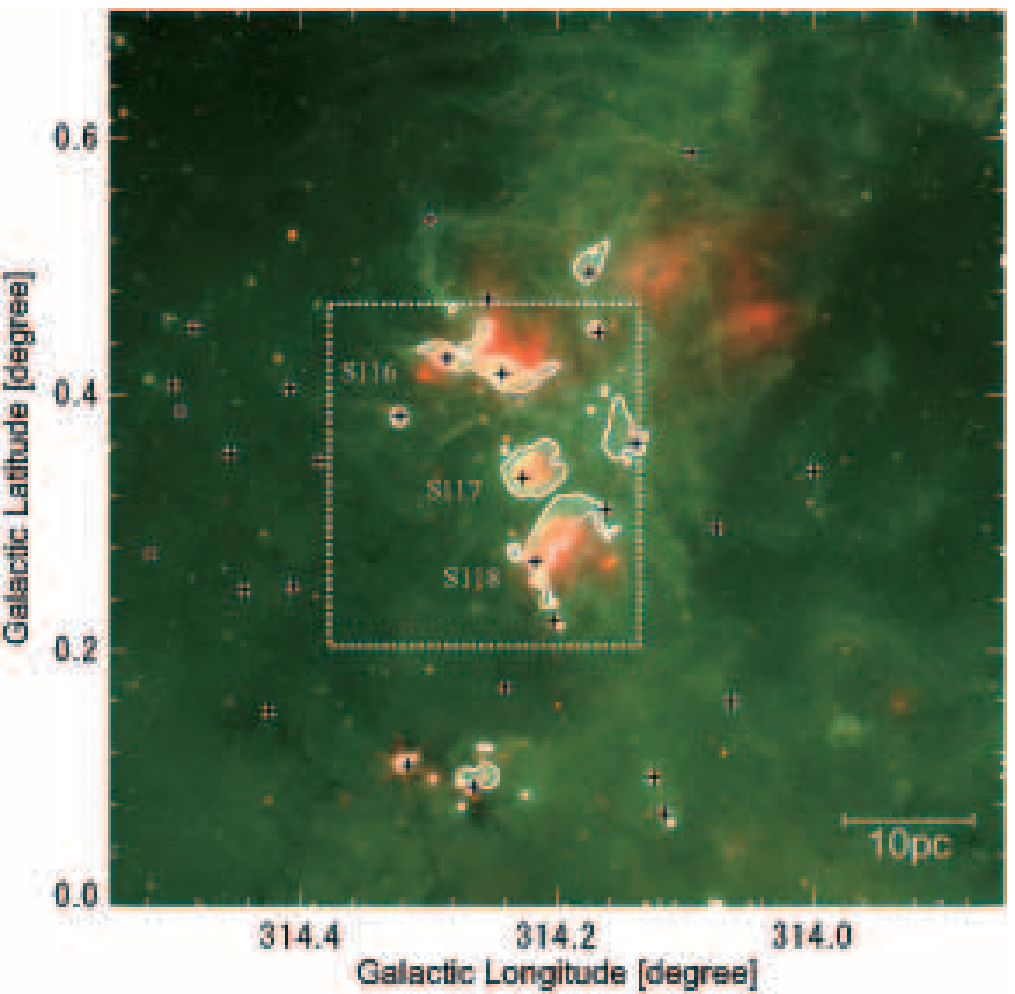} 
  \end{center}
\caption{The Spitzer image of the S116-S117-S118 region.
Green and red show the IRAC 8 \um \ emission \citep{ben03} and the MIPSGAL 24 \um \ emission \citep{car09}, respectively.
The crosses indicate YSOs catalogued by AKARI all-sky survey \citep{tot14}.
The white-dashed box shows the observed region with Mopra and ASTE. 
White contours show outlines of the 8 \um \ emission (100\,MJy\,str$^{-1}$) used in the following figures, where the 8 \um \ data are median-filtered. }
\label{fig:RGB}
\end{figure}

\begin{figure}
 \begin{center}
  \includegraphics[width=16cm]{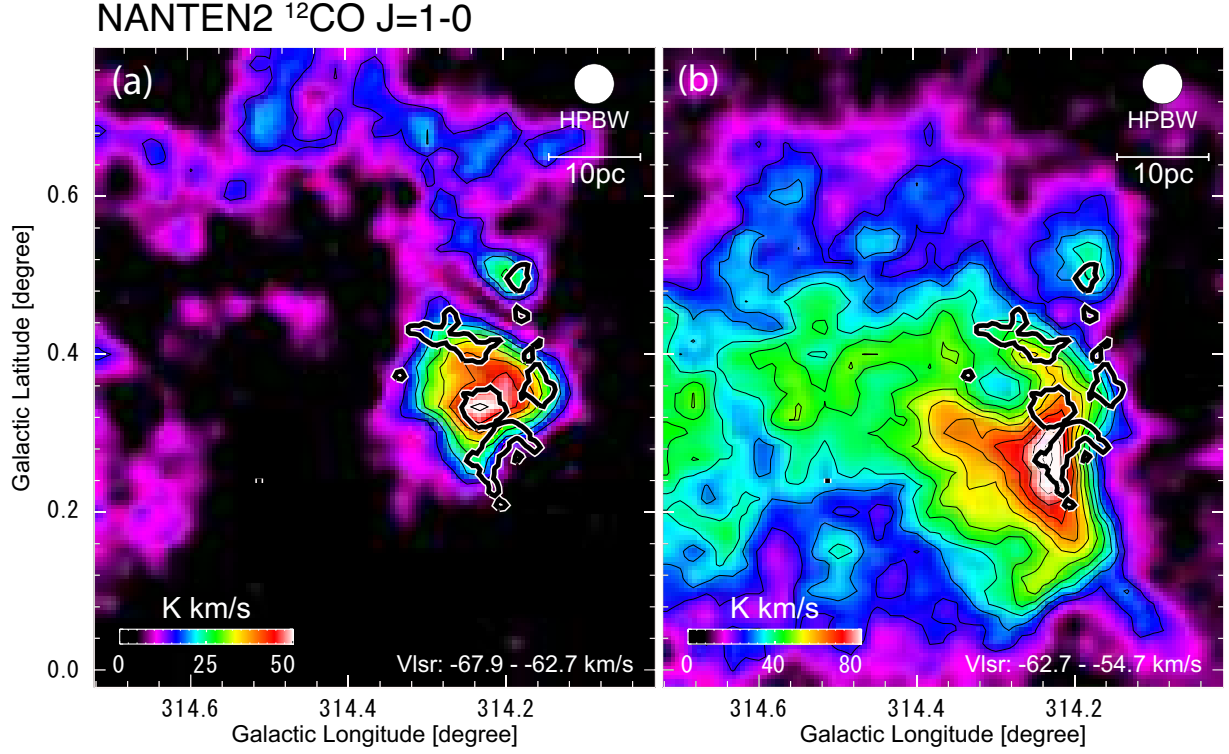} 
  \end{center}
\caption{(a) The \COa \ \Jeq 1--0 integrated intensity distribution obtained with NANTEN2 for the small cloud (a) and the large cloud (b). The bold contours indicate the Spitzer 8 \um \ outlines.}
\label{fig:nanten}
\end{figure}

\begin{figure}
 \begin{center}
  \includegraphics[width=16cm]{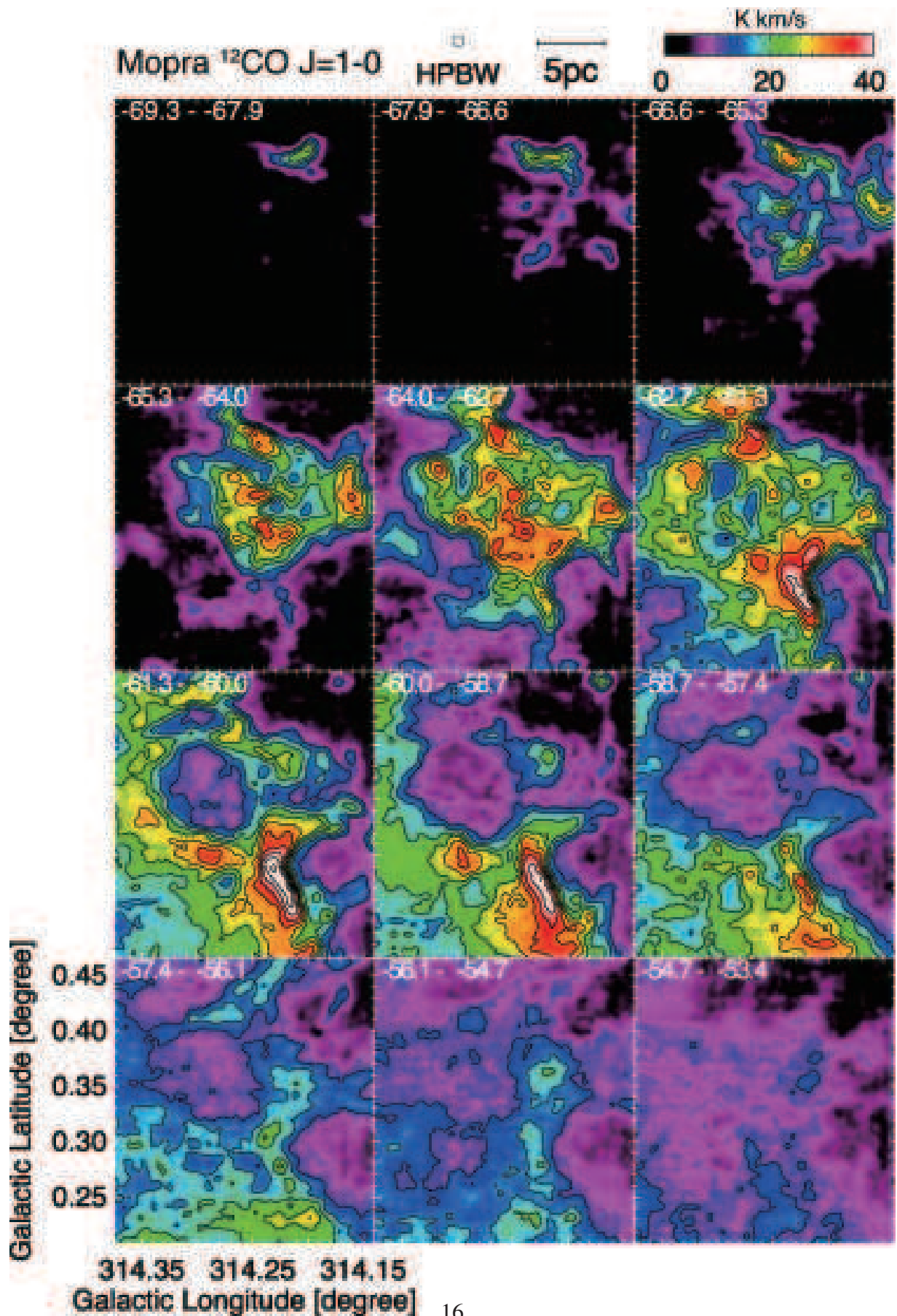} 
 \end{center}
\caption{\COa \ \Jeq 1--0 velocity-channel distributions every 1.3 \kms obtained with Mopra. The values printed in the top left corner of each panel denote the velocity range (km\,s$^{-1}$).}
  \label{fig:chmap}
\end{figure}

 \begin{figure}
  \begin{center}
 \includegraphics[width=16cm]{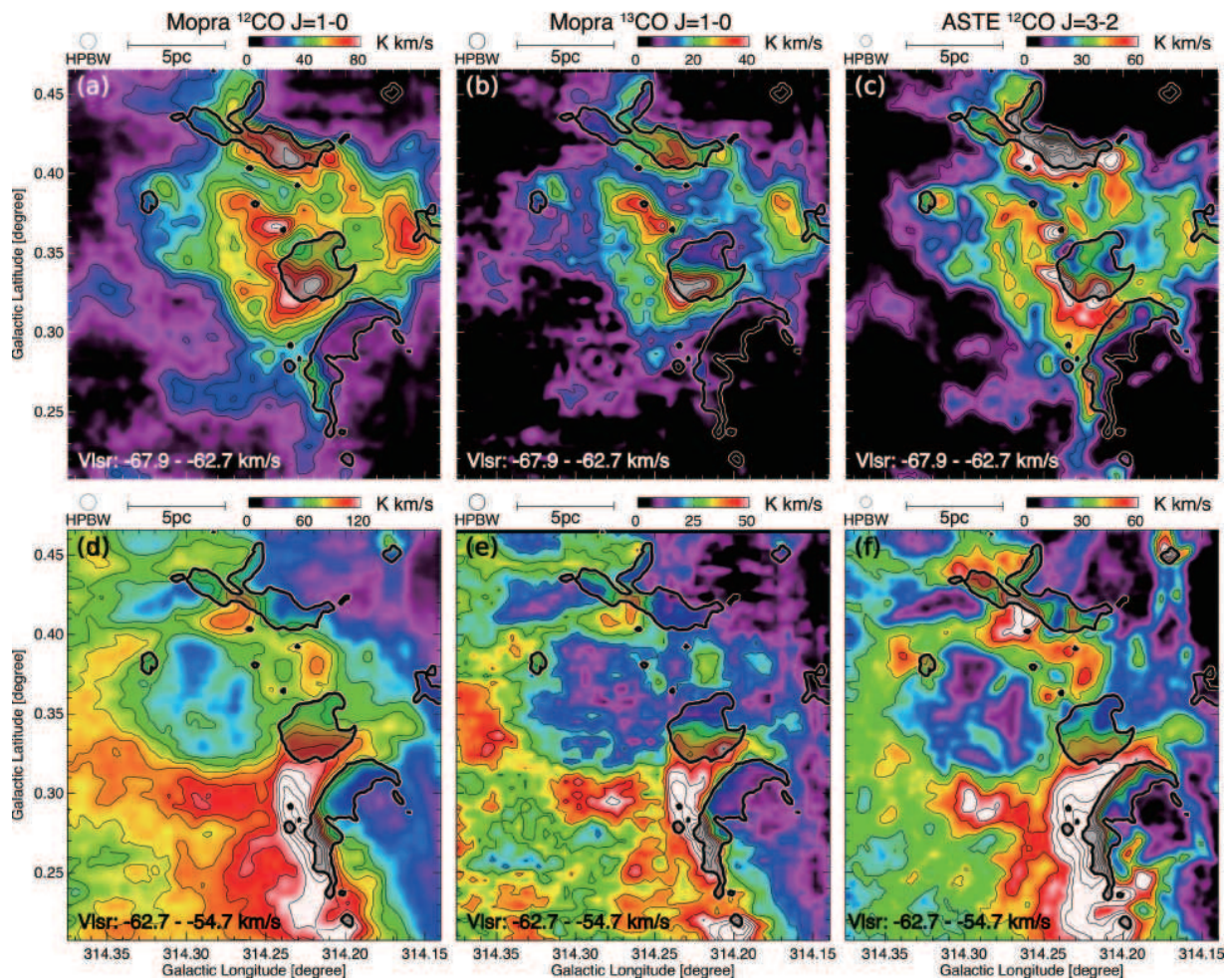}
  \end{center}
  \caption{
Distributions of the in \COa \ \Jeq 1--0, \COb \ \Jeq 1--0 and \COa \Jeq 3--2 transitions  for the small cloud ((a), (b), and (c)) and the large cloud ((d), (e), and (f))}.
The bold contours indicate the Spitzer 8 \um \ outlines.
  \label{fig:CO_3x2}
\end{figure}

\begin{figure}
 \begin{center}
  \includegraphics[width=16cm]{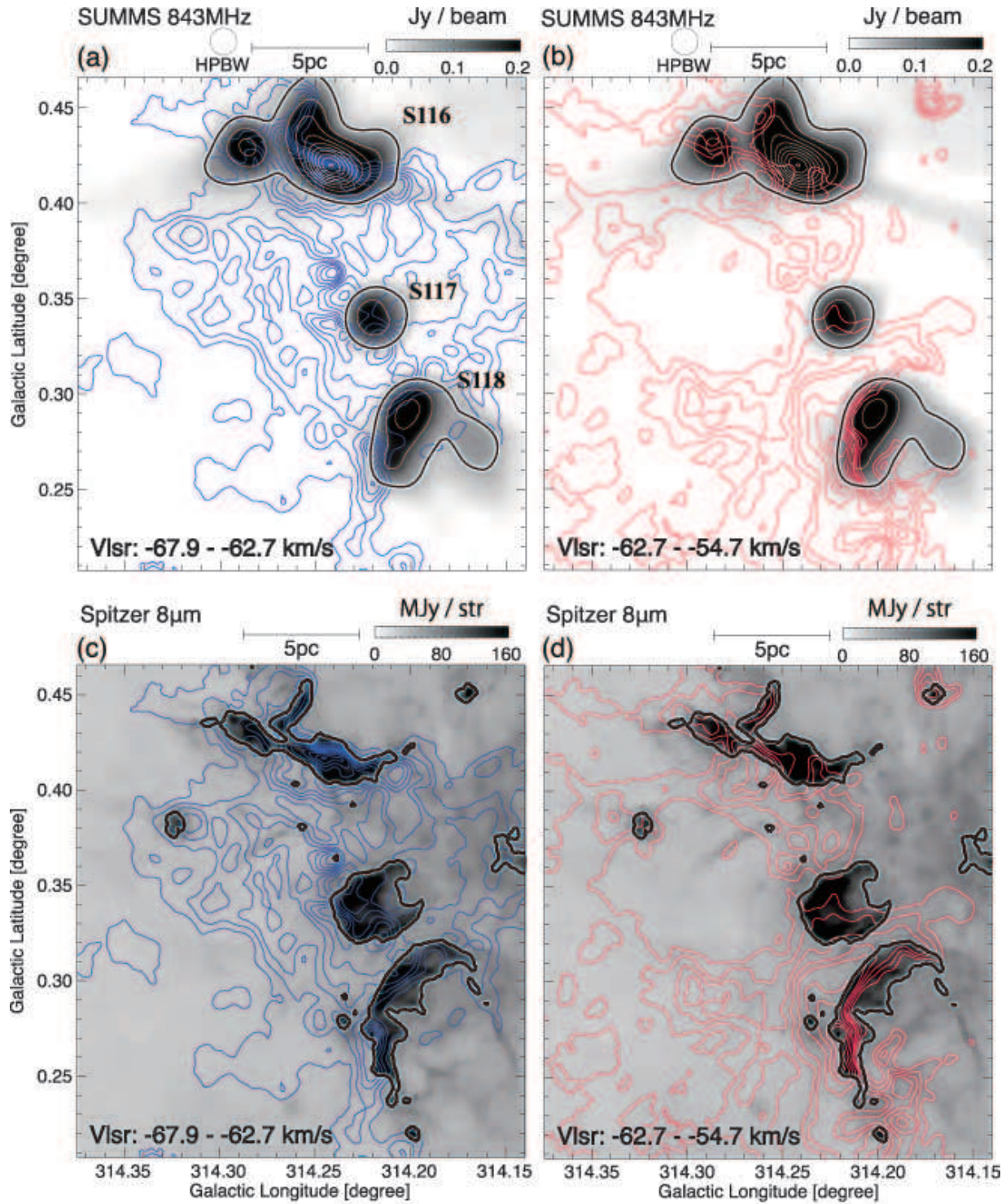} 
 \end{center}
\caption{\COa \ \Jeq 3--2 distributions superposed by the 843 MHz SUMSS radio continuum emission and the Spitzer 8 \um \ image; (a) the small cloud and the radio continuum image, (b) the large cloud and the radio continuum image, (c) the small cloud and the Spitzer 8 \um image, and (d) the large cloud and the Spitzer 8 \um image. }
\label{fig:radio}
\end{figure}

 \begin{figure}
  \begin{center}
 \includegraphics[width=16cm]{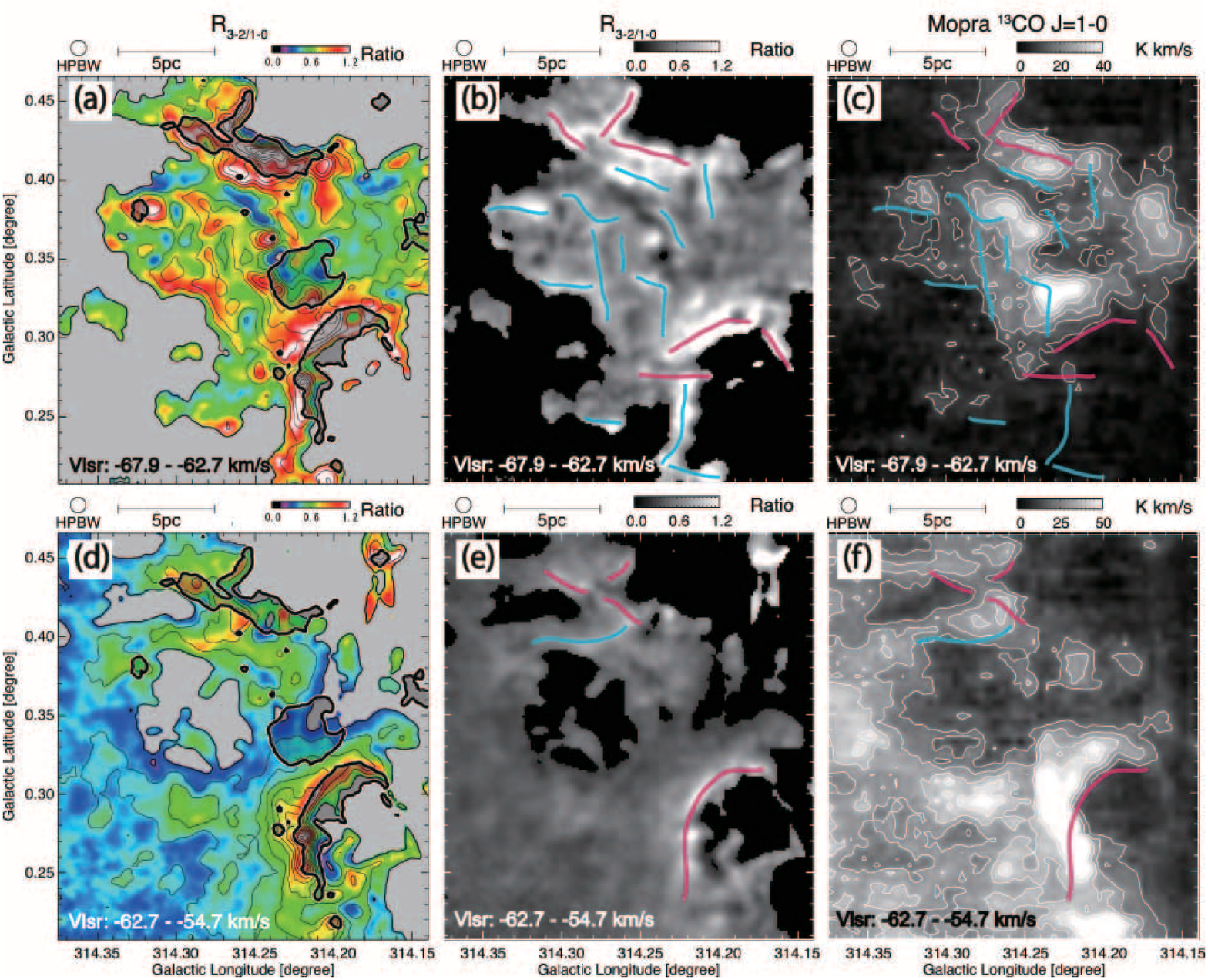}
  \end{center}
  \caption{
Distributions of the line intensity ratio between the \COa \ \Jeq 3--2 and \COa \ \Jeq 1--0 emission ($R_{3-2/1-0}$) for the small cloud ((a) and (b)) and the large cloud ((d) and (e)). (c) and (f) show the distribution of the \COb \ \Jeq 1--0 emission. Contours indicate the distribution of the \COa \ \Jeq 1--0 emission for (a) and (c), and the \COb \ \Jeq 1--0 emission for (c) and (f). The bold contours in (a) and (d) indicate the Spitzer 8 \um \ outlines. 
Red and blue lines represent the filamentary structures of $R_{3-2/1-0}$, where colors show that the filament is overlapping with the \HII \ regions (red), or not (blue). 
These filamentary features were identified by eye with a criteria of $R_{3-2/1-0}>0.9$ and length $>\sim2$\,pc.
 }
  \label{fig:ratio_3x2}
\end{figure}

\begin{figure}
 \begin{center}
  \includegraphics[width=16cm]{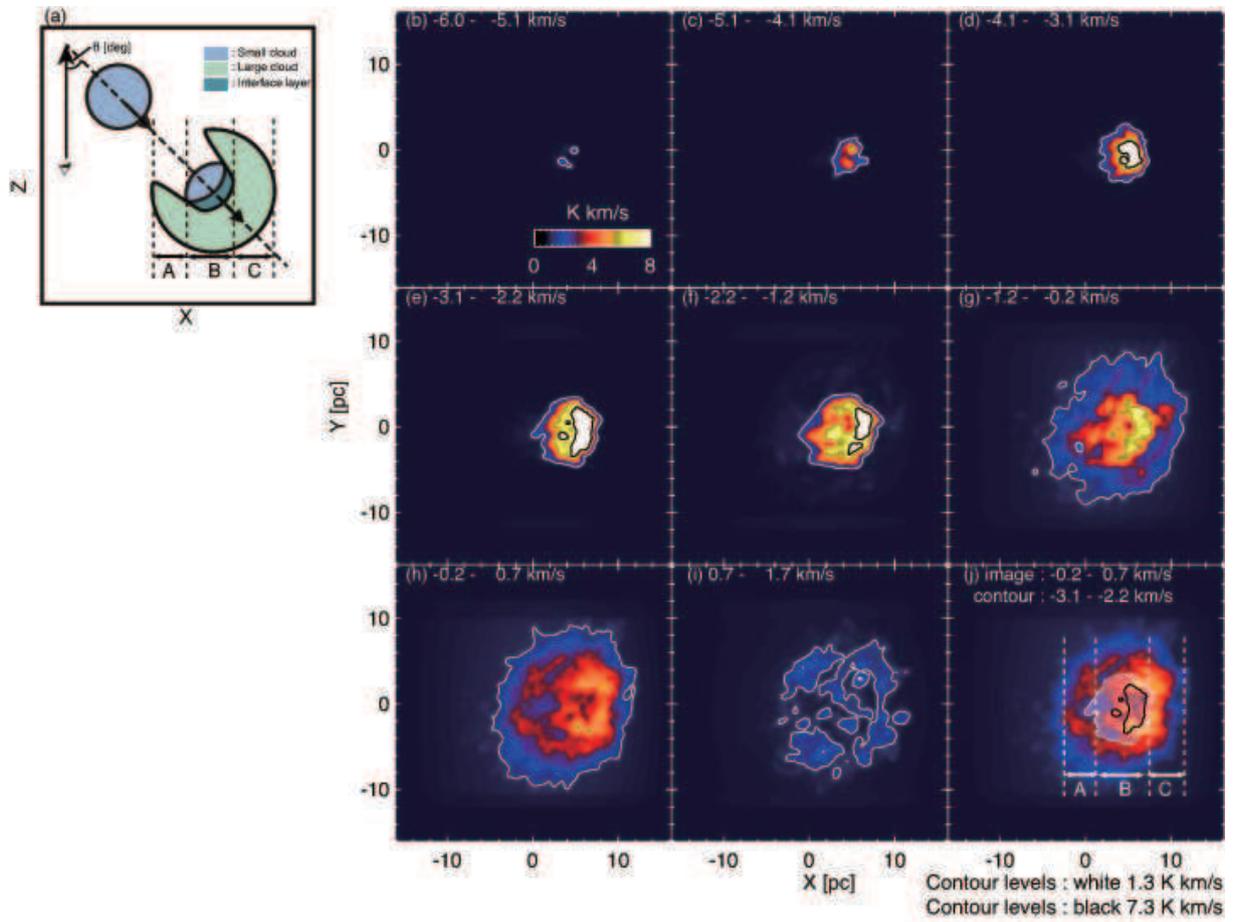} 
 \end{center}
\caption{
Synthetic observations of the numerical simulations of cloud-cloud collision by \citet{tak14}; (a) a schematic of the two clouds with an angle $\theta$ between the relative cloud motion and the line of sight, (b)--(i) velocity channel distributions every 0.9 km s$^{-1}$, and (j) an overlay of the small and large clouds (panels (e) and (h)).
}
  \label{fig:takahira_chmap}
\end{figure}

\begin{figure}
 \begin{center}
  \includegraphics[width=16cm]{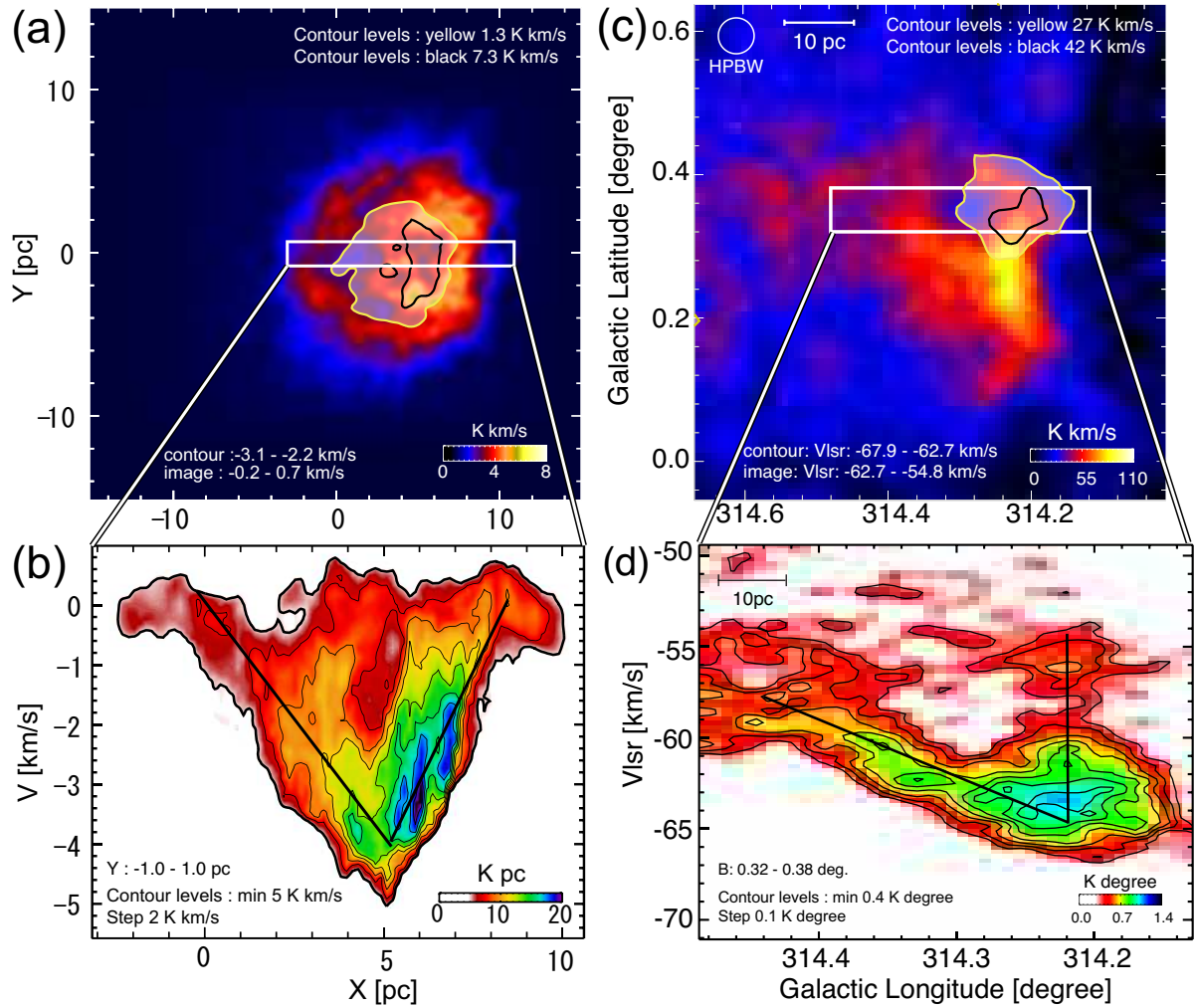} 
 \end{center}
\caption{Comparison between the synthetic observations and the observed CO distributions. (a) The spatial distribution of the small cloud and the large cloud of the synthetic observations. (b) Position-velocity diagram of the the synthetic observations. These images were reproduced from Takahira et al. (2014). (c) The $^{12}$CO $J =$ 1--0 spatial distribution obtained with NANTEN2 of the small cloud and the large cloud in S116-S117-S118. (d) Galactic longitude-velocity diagram in S116-S117-S118 of $^{12}$CO $J =$ 1--0 obtained with NANTEN2. }
\label{fig:pv}
\end{figure}

\begin{figure}
 \begin{center}
  \includegraphics[width=16cm]{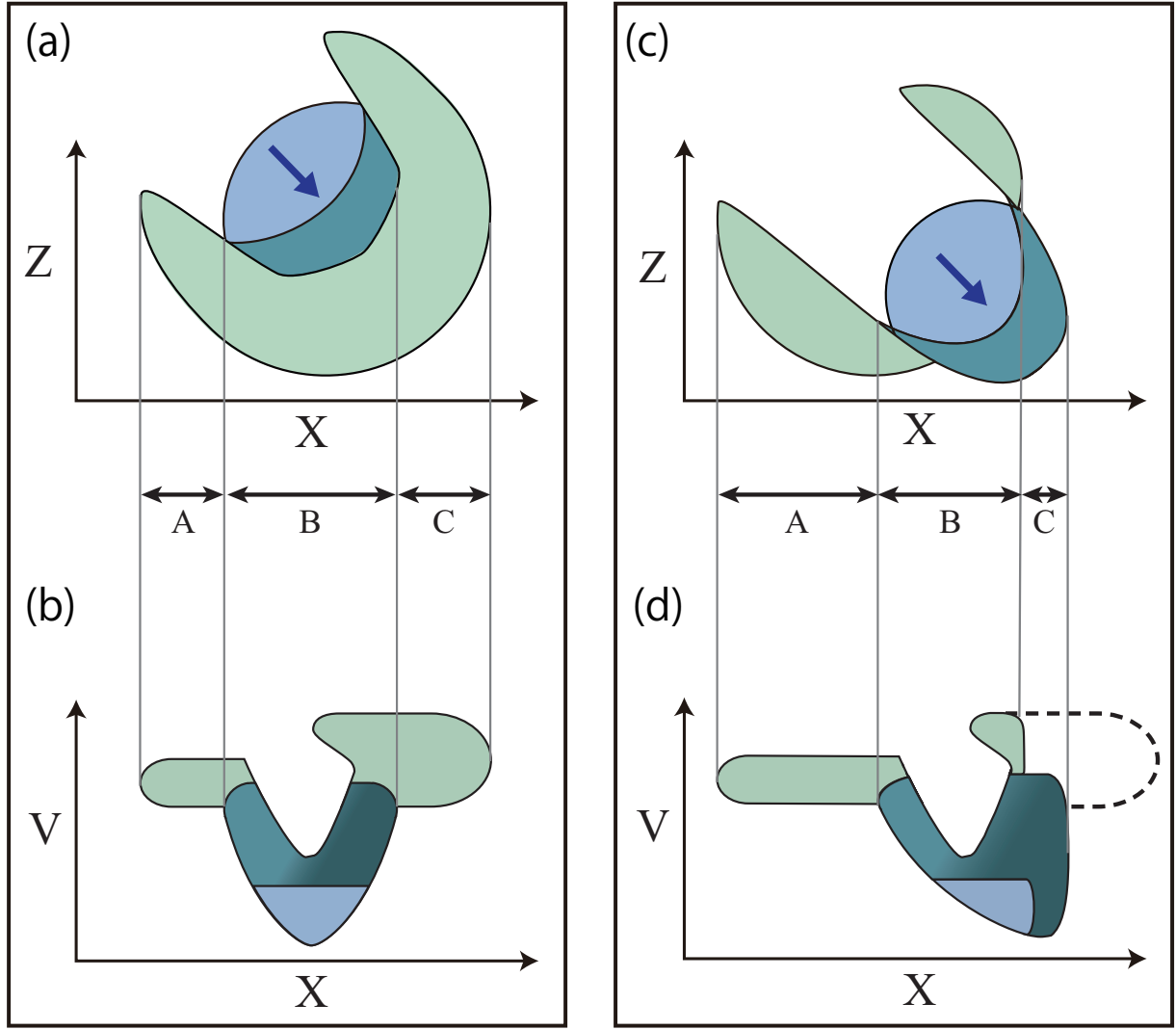} 
 \end{center}
\caption{ Schematic images of the two colliding clouds in the top views ((a) and (c)) and the position-velocity diagrams ((b) and (d)). 
X-axis and Z-axis are defined in the same way in Figure \ref{fig:takahira_chmap}, and V-axis corresponds to the radial velocity. (a) and (b) show the RCW120 case, and (c) and (d) show the S116-S117-S118 case. 
In (a) and (b), the two clouds can be divided into three sections as follows: A shows the large cloud and the cavity,  B shows the interface layer, the small cloud, and the large cloud with the cavity, and C shows the large cloud alone. 
In (c) and (d), the two clouds can be divided into three sections as following: A is the same as that of (a) and (b), B is the same as that of (a) and (b), and C shows the interface layer alone. 
The difference between the two models is whether the interface layer is inside or outside of the large cloud. 
In (d), there is no gas shown by the dashed line.
In section A, the majority of the clouds before the collision have been scrapped by the collision and are decreasing for the two models. 
In section B, the X-V diagrams show ''V-shape'' structure for the two models, although the right side of ''V'' in (b) is denser than that of (d).
In section C, there is only gas of the large cloud in (a) and (b), whereas there is only gas of the interface layer and it appears vertical in the X-V diagram in (c) and (d). 
}
\label{fig:schematic_pv}
\end{figure}

\begin{figure}
 \begin{center}
  \includegraphics[width=16cm]{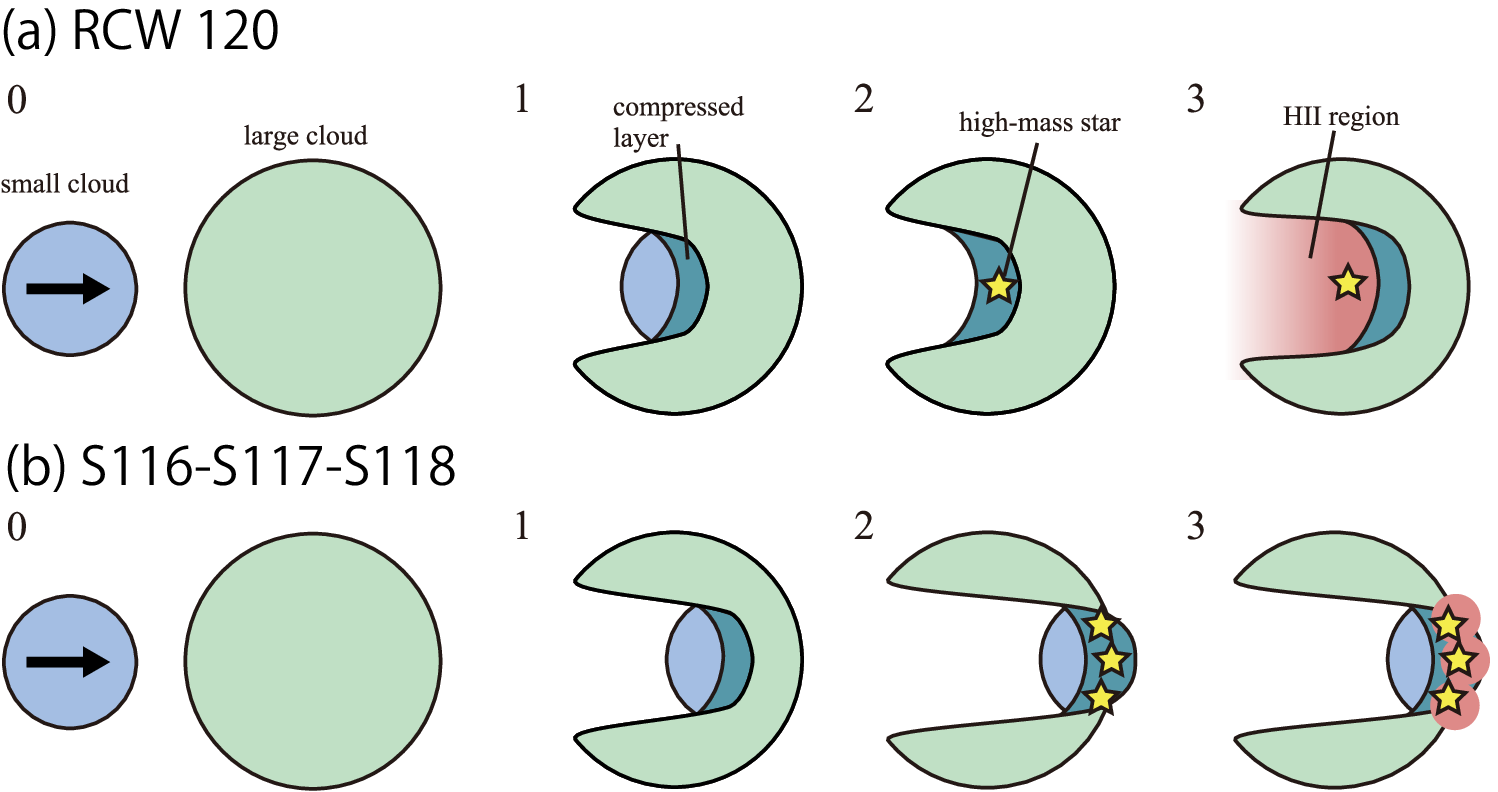} 
 \end{center}
\caption{
The schematic images of four phases in cloud-cloud collision in (a) RCW 120 and (b) S116--S117--S118. 
Phase 0 is prior to collision, Phase 1 the early phase, Phase 2 the intermediate phase when O star(s) formed, and Phase 3 the final phase when \HII \ regions (red color) formed.
The difference between the two cases (a) and (b) is the location of the O star(s). 
In RCW120 the O star is inside the cavity ((a) phase 3), whereas in S116--S117--S118 the O stars are located outside the cavity ((b) phase 3). 
Unlike RCW120, no \HII \ region is formed inside the cavity in S116--S117--S118. 
The O star formation in S116--S117--S118 happened close to the edge of the large cloud, making the interface layer exposed to the outside of the large cloud (see text).
}
\label{fig:schematic_star}
\end{figure}

%% file: s9_tables.tex
\begin{table}
\tbl{Physical properties of the Spitzer bubbles}{%
\begin{tabular}{lccc}
\hline\noalign{\vskip3pt} 
Spitzer bubble Name & $\log N_{\rm L}$ & Sp. type & $M_*$ \\  
 & (photons) & & \Msun \\
\hline\noalign{\vskip3pt} 
S116 & 48.9 & O6--O6.5 & 30 \\
S117 & 48.0 & O9 & 20 \\
S118 & 48.5 & O7.5--O8 & 23 \\
\hline\noalign{\vskip3pt}
\end{tabular}}
\label{table:1}
\begin{tabnote}
Notes. Column 1: name of the Spitzer bubble. Column 2: number of UV photons estimated from the flux of radio continuum of SUMMS 843 MHz survey \citep{boc99, mau03} by using relationship described in \citep{kur94}. Column 3: estimated spectral type of the exciting star in the region by using $N_{\rm L}$--$T_{\rm eff}$ relationship \citep{pan73}. Column 4: estimated stellar mass.
\end{tabnote}
\end{table}

\begin{table}[h] 
\begin{center}
\tbl{The initial conditions of the numerical simulations (Takahira et al. 2014) }{%
\begin{tabular}{@{}cccccc@{}} \noalign{\vskip3pt}
\hline\hline
\multicolumn{1}{c}{Box size [pc]} & $30 \times 30 \times 30$  &  &  &  \\ [2pt]
\noalign{\vskip3pt}
Resolution [pc] & 0.06 &  &  \\
Collsion velocity [km s$^{-1}$] & 10 (7)$^{\dag}$ & & \\
\hline
Parameter & The small cloud & The large cloud & note    \\
\hline
Temperature [K] & 120 & 240 &  \\
Free-fall time [Myr] & 5.31 & 7.29 &  \\
Radius [pc] & 3.5 & 7.2 &  \\
Mass [$M_{\odot}$] & 417 & 1635 &  \\
Velocity dispersion [km s$^{-1}$] & 1.25 & 1.71 &  \\
Average Density [cm $^{-3}$] & 47.4 & 25.3 & Assumed a Bonner-Ebert sphere  \\
\hline\noalign{\vskip3pt} 
\end{tabular}} 
\label{tab:first} 
\begin{tabnote}
{\hbox to 0pt{\parbox{150mm}{\footnotesize
Note. \footnotemark[$\dag$] The initial relative velocity between the two clouds is set to 10 km s$^{-1}$, whereas the collisional interaction decelerates the relative velocity to about 7 km s$^{-1}$ in 1.6 Myrs after the onset of the collision. The present synthetic observations are made for a relative velocity 7 km s$^{-1}$ at 1.6 Myrs.
\par\noindent
\phantom{0}
\par
\hangindent6pt\noindent
}\hss}}
\end{tabnote} 
\end{center} 
\end{table}